# Hexagonal germanium grown by molecular beam epitaxy on self-assisted GaAs nanowires

I. Dudko[1,2,3], T. Dursap[1], A. D. Lamirand[1], C. Botella[1], P. Regreny[1], A. Danescu[1], S. Brottet[1], M. Bugnet[2], S. Walia[2,3], N. Chauvin[1], J. Penuelas[1]

[1] Univ Lyon, Ecole Centrale de Lyon, CNRS, INSA Lyon, Université Claude Bernard Lyon 1, CPE Lyon, INL, UMR 5270, 69130 Ecully, France

[2] School of Engineering, RMIT University, Melbourne 3001, Victoria, Australia

[3] Functional Materials and Microsystems, Research Group and Micro Nano Research Facility, RMIT University, Melbourne 3001, Victoria, Australia

[4] Univ Lyon, CNRS, INSA Lyon, UCBL, MATEIS, UMR 5510, 69621 Villeurbanne, France

[*] To whom correspondence should be addressed. E-mail : jose.penuelas@ec-lyon.fr

**Abstract**: Hexagonal group IV materials like silicon and germanium are expected to display remarkable optoelectronic properties for future development of photonic technologies. However, the fabrication of hexagonal group IV semiconductors within the vapour-liquid-solid method has been obtained using gold as a catalyst thus far. In this letter, we show the synthesis of hexagonal Ge on self-assisted GaAs nanowires using molecular beam epitaxy. With an accurate tuning of the Ga and As molecular beam flux we selected the crystal phase, cubic or hexagonal, of the GaAs NWs during the growth. A 500 nm-long hexagonal segment of Ge with high structural quality and without any visible defects is obtained, and we show that germanium keeps the crystal phase of the core using scanning transmission electron microscopy. Finally X-ray Photoelectron Spectroscopy reveals a strong incorporation of As in the Ge. This study demonstrates the first growth of hexagonal Ge in the Au-free approach, integrated on silicon substrate.

The recent experimental demonstration of direct band gap in hexagonal Ge (*h*-Ge) and SiGe alloy [1] has opened new opportunities in the field of silicon photonics [2]–[4]. While in its natural cubic phase Ge has an indirect band-gap, the hexagonal phase was predicted to have a direct band gap [5]–[9]. The hexagonal phases of Si and Ge were achieved in 2015 [10] and 2017 [11], respectively, thanks to the crystal transfer method. This method consists in the growth of III-V nanowires (NWs) with wurtzite (WZ) crystal structure, and the subsequent epitaxial growth of the group IV semiconductor (Si, Ge or an alloy) on the NW facets. Due to symmetry and lattice mismatch considerations, the group IV semiconductor can adopt the hexagonal phase in a core / shell configuration. Compared to *h*-Ge grown by metalorganic chemical vapor deposition (MOCVD) [1] molecular beam epitaxy (MBE) provides distinct advantages such as a high purity of the grown materials and an accurate control of the thickness down to the atomic monolayer, as well as the growth far from thermodynamic equilibrium, which are relevant to fabricate optimized quantum heterostructures for instance. Moreover, III-V MBE reactors can be easily equipped with Ge and/or Si evaporators.

In the pioneering work of Fadaly *et al.* [1] III-V NWs were grown on GaAs substrate with Au catalyst using MOCVD. While Au is probably the most popular catalyst for the NW growth using the vapor liquid solid (VLS) method, Au is known to induce defects and can diffuse into the semiconductor [12], [13]. An alternative solution consists in using the self-assisted method, where Au is replaced by Ga for the growth of GaAs NWs [14], [15]. Although replacing Au by Ga is promising to avoid the formation of defects, it was observed that self-assisted GaAs NWs exhibit mostly the zinc-blende (ZB) phase, which is detrimental to the fabrication of *h*-Ge. The occurrence of two crystal phases in GaAs NWs (WZ for Au catalyst and ZB for Ga catalyst) is closely related to the value of contact angle of the liquid catalyst during the VLS growth [16]–[18], and various strategies have been tested in order to control the synthesis of WZ in NW geometry [19], [20]. Recently, we demonstrated the growth of a long segment of WZ GaAs using the self-assisted method by an accurate control of the group III [21] or group V flux [22] to ultimately reach a stationary state of the catalyst contact angle in the range inducing the WZ phase [22]. The facets of obtained WZ GaAs

NWs thus provide an ideal template for the epitaxial growth of *h*-Ge, in addition to their low lattice mismatch.

In this work, we demonstrate the growth of epitaxial *h*-Ge on the facets of self-assisted GaAs NWs by MBE. Our technique is gold catalyst-free and is realized directly on Si substrate. The as-grown NWs are thoroughly characterized using microscopic and spectroscopic techniques, and highlight the single-crystal quality of the germanium shell. This Au-free approach opens new pathways for integrating *h*-Ge NWs in photonic devices such as lasers or near-infrared detectors **[23]**.

GaAs/Ge core-shell NWs were epitaxially grown on Si (111) substrate by MBE using the VLS mechanism. Before the introduction to the reactor, each Si substrate was cleaned with acetone and ethanol in an ultrasonic bath for 5 min, and degassed at 200°C in ultra-high vacuum. Afterwards, the substrate was heated to 450°C and 1 monolayer (ML) of Ga was pre-deposited during 2 seconds at 0.5 ML.s$^{-1}$, quoted in units of equivalent growth rates of GaAs 2D layers measured by RHEED oscillations on a GaAs substrate **[27]**. The substrate temperature was increased to 580°C for the growth of GaAs core. In order to obtain a long GaAs WZ segment as described in Figure 1(a) the crystal phase was tuned by a modification of the V/III ratio during the VLS growth. GaAs NWs were grown during 20 min with Ga and As$_4$ fluxes of 0.5 ML s$^{-1}$ and 1.2 ML s$^{-1}$, respectively, corresponding to a V/III flux ratio 2.4 in order to obtain the ZB phase. To achieve the WZ phase, the V/III ratio was increased to 4.3 for 20 minutes (As$_4$ flux was increased to 2.1 ML s$^{-1}$). The growth of GaAs core was terminated by closing the Ga shutter in order to consume the droplet. The growth was continuously monitored by *in situ* reflection high energy electron diffraction (RHEED), which characterizes the crystal structure of the samples during the growth, to obtain real time information of the growing crystal phase. Figure 1(b) and 1(c) illustrate the growing WZ phase of GaAs during 20 minutes, with the 4.3 V/III ratio. The intensity of the WZ spots on the RHEED increased after the 10$^{th}$ minute of growth (Figure 1(b)) and the progressive intensity of WZ signal proved the growth of long hexagonal segment. The Ge shell was then epitaxially grown around the self-assisted GaAs core NWs. The Ge shell was grown at 400°C for 30 min, which corresponds to a planar growth rate of 0.15 ML/s. RHEED images in Figure 1(d) and Figure 1(e) were recorded during the Ge growth after the 3$^{rd}$ and 10$^{th}$ minute, respectively. The RHEED signal did not change from Figure 1(c) to Figure 1(d) confirming that Ge followed the crystal structure of GaAs core. Figure 1(f) shows the scanning electron microscopy (SEM) image of the NWs, their length is about 4.5 µm and their diameter about 115 nm. Thanks to this growth methodology, we expect NWs having a long hexagonal segment located near the tip as described in figure 1(a).

High-resolution scanning transmission electron microscopy (STEM) imaging in high angle annular dark field (HAADF) mode and elemental mapping using electron energy-loss spectroscopy (EELS) were performed to investigate the crystal structure and the epitaxial interface between GaAs and Ge. Several NWs were analyzed with the same results. The top part of the NWs can be easily identified from the Ge enrichment at the NW tip (see Figure S3 in the supplementary information section), which is a consequence of the Ge evaporator orientation (5° of incidence to the normal of the substrate). In agreement with the evolution of the RHEED pattern during the growth, cubic Ge is observed on ZB GaAs (see Figure S1 in the supplementary information section), while *h*-Ge is observed on WZ GaAs. As expected, the upper part of the NW exhibit a *h*-Ge / WZ GaAs segment of 500 nm. Figure 2(b) shows a high-resolution STEM-HAADF image of this extended hexagonal segment. Interestingly, no extended defects **[24]** were identified, the low lattice mismatch between Ge and GaAs is probably accommodated by surface relaxation. Chemical mapping using EELS shown in Figure 2(b) highlights a Ge shell thickness of about 6 nm. The clean interface seems abrupt, and exhibits no Ga diffusion into the Ge shell.

X-ray photoelectron spectroscopy (XPS) was used to investigate the presence of impurities and possible As contamination in the Ge shell. XPS measurements were carried out in a Prevac spectrometer equipped with a focused monochromatic X-ray source Al Kα (1486.6 eV), in a base pressure less than $10^{-9}$ mbar. The different spectra were recorded at 45° exit angle to maximize the signal coming from the side of the NWs. Following the method described previously, a thicker Ge shell was grown on GaAs NWs with a hexagonal segment estimated to 500 nm. The shell thickness was increased to 20 nm in order to maximize the signal from the Ge shell. After the growth, the sample was transferred in ultra-high vacuum to the XPS chamber, which is directly connected to the MBE reactor, for further analysis.

A large survey (shown in Figure 3(a)), O 1$s$, As 3$d$, Ge 3$d$ and Ga 3$d$ core level spectra of the sample were recorded and compared to a Ge substrate reference. The position and shape of Ge peaks from the NWs reproduce exactly the Ge peaks from the reference substrate, thus confirming the presence of Ge-Ge bonds. The Ge 3$d_{5/2,3/2}$ multiplet, shown in Figure 3(b), was fitted using a spin-orbit splitting of 0.585 eV and an intensity ratio of 0.67, as expected statistically for $d$ electrons and in agreement with literature [25]. The inelastic electron background was taken into account via a Shirley background correction. Ge 3$d$ was found at 29.9 eV, corresponding to a slightly higher value than literature [26].

The comparison between the two surveys reveals very weak signals coming from Ga 2$p$ core level at about 1117 eV and from As Auger electrons around about 260 eV. Both are indicated in Figure 3(a). The amount of As, Ge and Ga at the extreme surface of the NWs can be easily compared in the Figure 3(b). The Ga 3$d$ was barely visible in contrast to As 3$d$, suggesting a slight diffusion or contamination of As in Ge. By considering a homogeneous As$_y$Ge overlayer in the NW, y is evaluated around 0.076 from the ratio of the 3$d$ fitted peak areas times their cross sections. This amount might be overestimated as the fit allows to disentangle electrons from Ge metal loss of electrons and from As 3$d$ core level, but does not to take into account the Shirley background. The presence of As in the Ge shell was already reported in ref [1] and attributed to As diffusion from the NW core due to high growth temperature (about 650°C). However, in our case the growth was performed at only 400°C, suggesting possible As contamination from the residual atmosphere in the MBE chamber during the germanium growth.

In conclusion, we demonstrated the growth of hexagonal Ge on self-assisted wurtzite GaAs segments in NW geometry. The core-shell NWs were grown directly on Si substrate using MBE. We succeeded to grow a long hexagonal segment of Ge with high structural quality that was confirmed by STEM. The XPS analysis reports the contamination of As in Ge, which strongly modify the electronic properties of $h$-Ge. Our results open the way to the fabrication of gold-free quantum heterostructures based on hexagonal Ge and the realization of associated devices.

The data that support the findings of this study are available from the corresponding author upon reasonable request.

The authors thank the NanoLyon platform for access to the equipments and J. B. Goure for technical assistance. The authors acknowledge the French Agence Nationale de la Recherche (ANR) for funding under project BEEP (ANR-18-CE05-0017-01). This work was supported by the LABEX iMUST (ANR-10-LABX-0064) of Université de Lyon, within the program "Investissements d'Avenir" (ANR-11-IDEX-0007) operated by the French National Research Agency (ANR). The STEM work was performed at the consortium Lyon-St-Etienne de microscopie. The authors are grateful to Y. Lefkir and S. Reynaud for technical assistance using the Jeol NeoARM instrument. This project has received funding from the European Union's Horizon 2020 research and innovation programme under the Marie Skłodowska-Curie grant agreement No 801512.

**Figures**

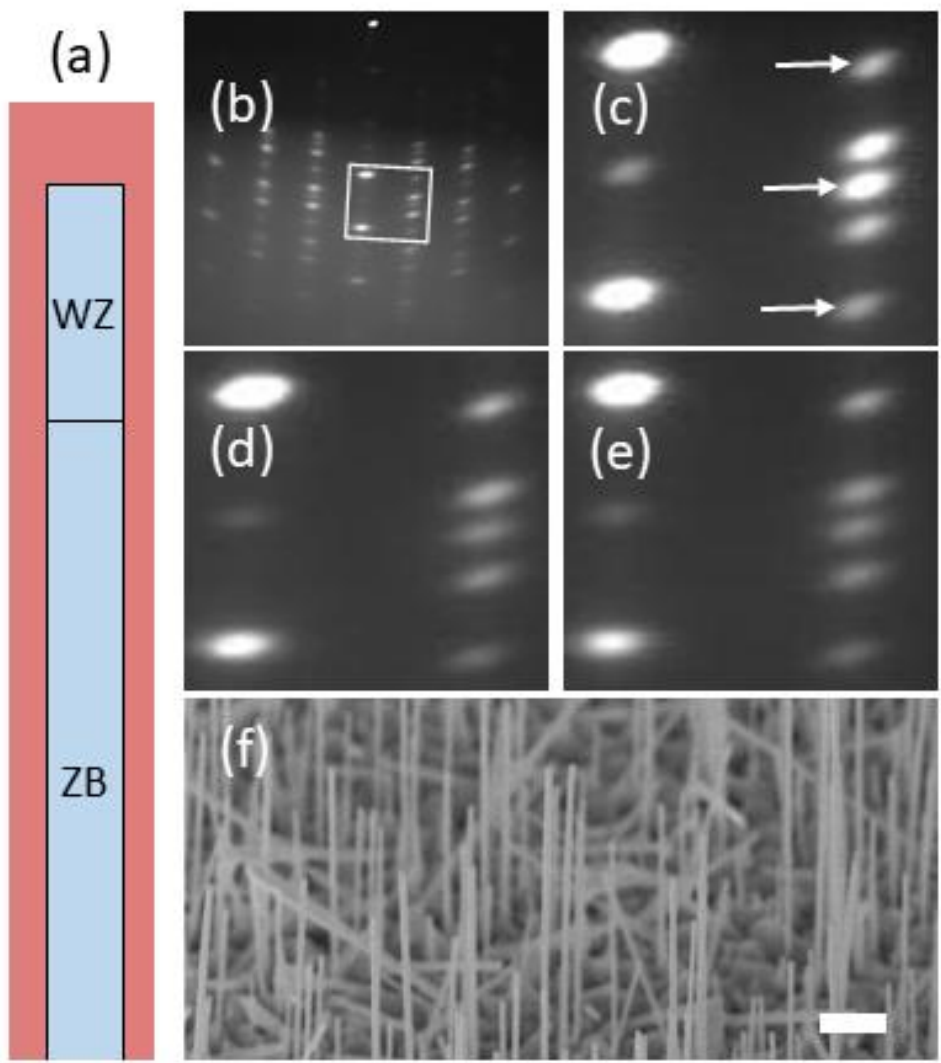

**Figure 1:** (a) Schematic of a core shell NW with controlled crystal phase. The samples were monitored *in situ* by RHEED to obtain real time information on the crystal structure of GaAs/Ge NWs. RHEED patterns recorded along the [1-10] azimuth during the growth of WZ GaAs at (b) 29 min (c) 35 min and of hexagonal Ge at (d) 3 min (e) 10 min. The WZ spots are highlighted with the white arrows. (f) 45° tilt SEM image (secondary electron contrast) showing the GaAs/Ge NWs. The scale bar is 1 μm.

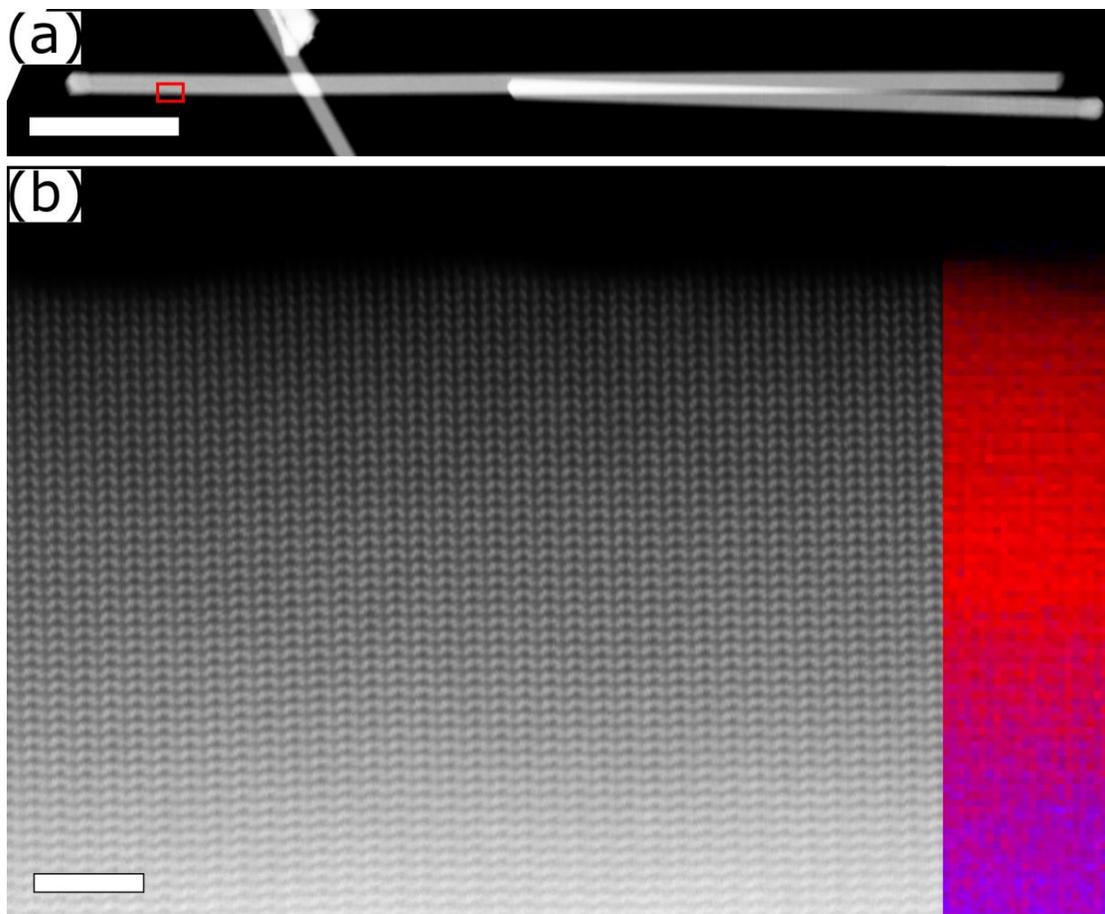

**Figure 2:** (a) STEM-HAADF overview of a NW. Scale bar is 500 nm. (b) High resolution STEM-HAADF image of the WZ structure with a superimposed Ga (blue) and Ge (red) maps extracted from multiple linear least square (MLLS) fitting. Scale bar is 2 nm. The region shown in (b) is highlighted by a red rectangle in (a). (c) Superimposed maps of the Ga core (blue) and the Ge shell (red) of a NW cross section. Scale bar is 50 nm. The STEM-HAADF and STEM-EELS experiments were carried out in a Jeol ARM 200, equipped with a high brightness cold-FEG, a last generation CEOS aberration-corrector (ASCOR) of the condenser lenses, a Gatan GIF Quantum, and operated at 200 kV.

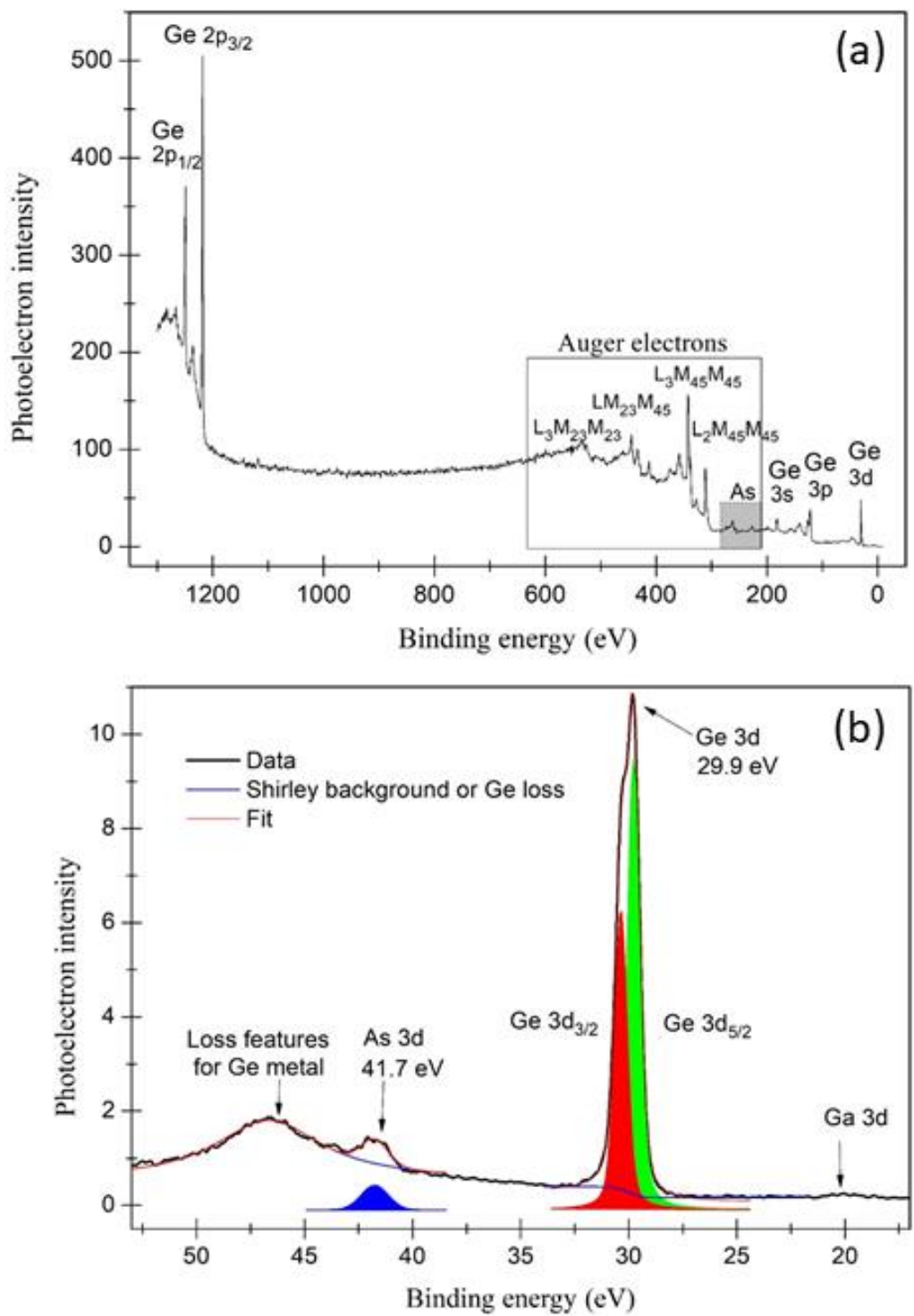

**Figure 3**: XPS spectra of GaAs/Ge nanowires (a) full range survey spectrum (b) Ge 3$d$ core level spectrum. The energy calibration was checked with gold reference (Au 4f at 84(+/-0.1) eV) just before the experiment. A Ge substrate was used as a reference sample. The C1s signal is found negligible for all samples. There is no evidence for oxygen in the NWs, as expected from a growth and transfer under UHV.

# Supporting Information for:

# Hexagonal germanium grown by molecular beam epitaxy on self-assisted GaAs nanowires.

*I. Dudko[1,2,3], T. Dursap[1], A. D. Lamirand[1], C. Botella[1], P. Regreny[1], A. Danescu[1], S. Brottet[1], M. Bugnet[2], S. Walia[2,3], N. Chauvin[1], J. Penuelas[1]*

[1] Univ Lyon, Ecole Centrale de Lyon, CNRS, INSA Lyon, Université Claude Bernard Lyon 1, CPE Lyon, INL, UMR 5270, 69130 Ecully, France

[2] School of Engineering, RMIT University, Melbourne 3001, Victoria, Australia

[3] Functional Materials and Microsystems, Research Group and Micro Nano Research Facility, RMIT University, Melbourne 3001, Victoria, Australia

[4] Univ Lyon, CNRS, INSA Lyon, UCBL, MATEIS, UMR 5510, 69621 Villeurbanne, France

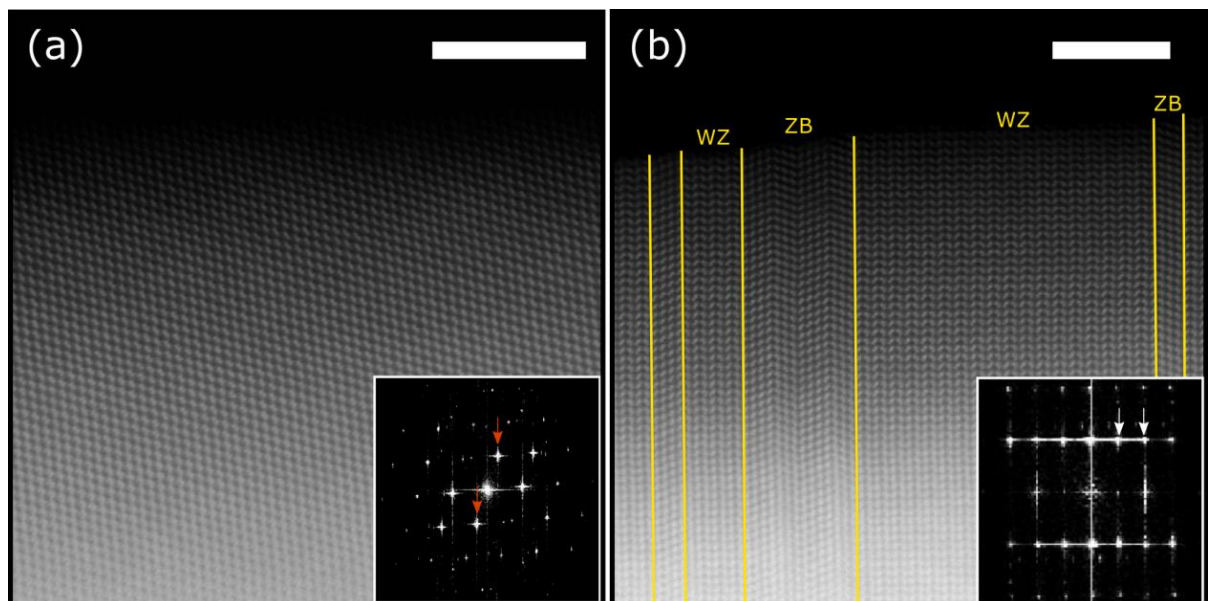

**Figure S1:** High-resolution STEM-HAADF images of the NW showing (a) the ZB crystalline structure and (b) a mix of the ZB and WZ phase. Each structures are delimited and identified in yellow. The insets show the FFT of the ZB phase and the WZ part of (a) and (b), respectively. The Red and white arrows highlight the ZB and WZ spots, respectively. Scale bars are 5 nm.

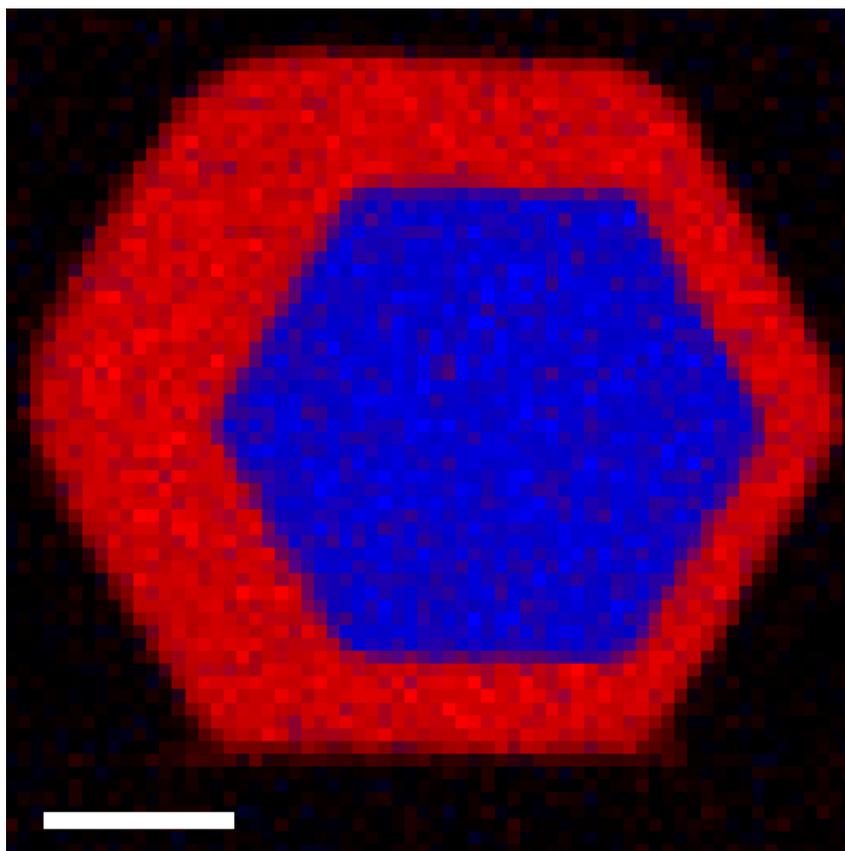

**Figure S2:** Superimposed maps of the Ga core (blue) and the Ge shell (red) of a NW cross section. Scale bar is 50 nm.

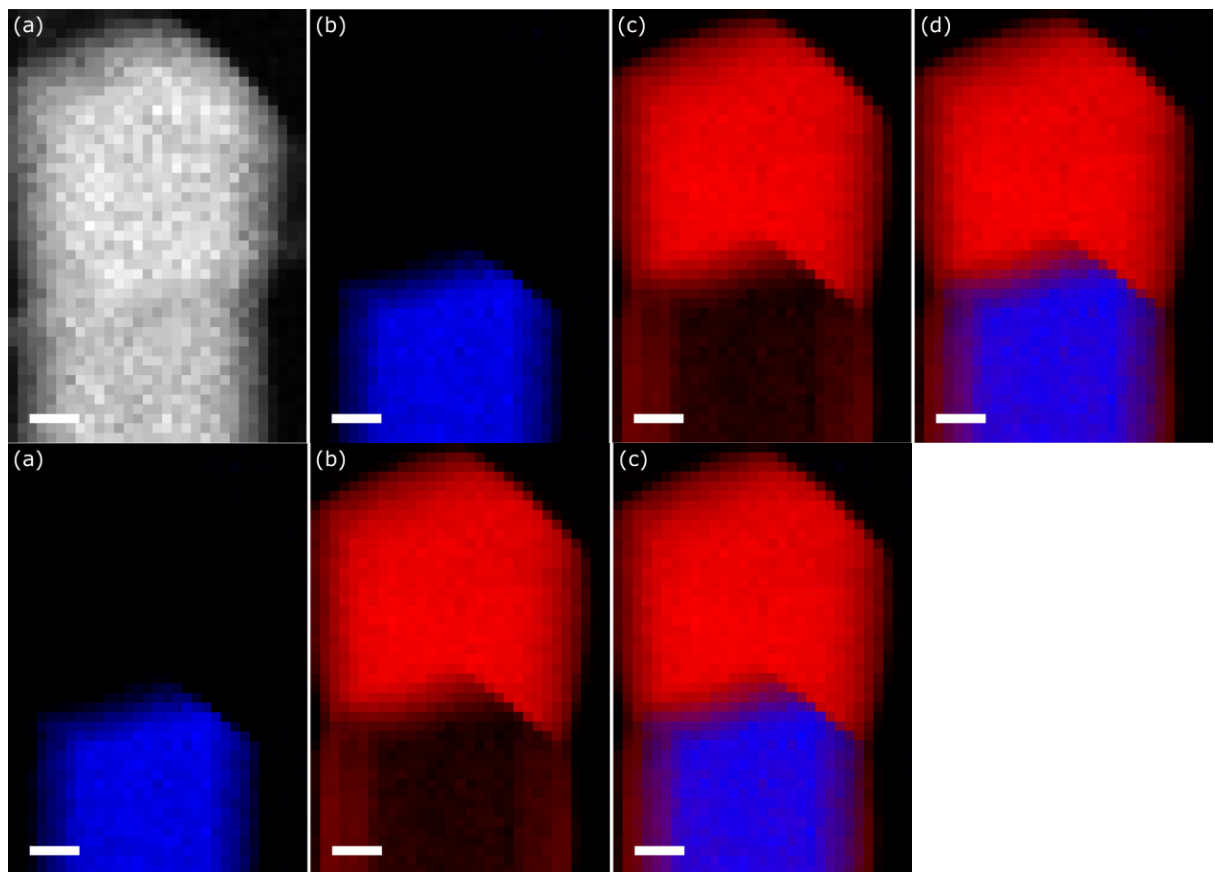

**Figure S3:** (a) HAADF-STEM image showing the top of the NW. (b) Ga core, (c) Ge shell, and (d) superimposed maps of the Ga core (blue) and Ge shell (red) extracted from multiple linear least square fitting. Scale bars are 20 nm.